\documentclass{emulateapj}
\usepackage{graphicx}
\usepackage{natbib}

\singlespace
\shortauthors{Kogut et al.}
\shorttitle{WMAP Foreground Polarization}
\slugcomment{Submitted to The Astrophysical Journal}

\begin{document}

% ---------------------Title ---------------------
\title{Three-Year Wilkinson Microwave Anisotropy Probe 
({\sl WMAP}\altaffilmark{1}) Observations: \\
Foreground Polarization}

% --------------------- Author list ---------------------
\author{A. Kogut\altaffilmark{2}, 
J. Dunkley\altaffilmark{3},
C. L. Bennett\altaffilmark{4},
O. Dor\'e\altaffilmark{5,6},
B. Gold\altaffilmark{4},
M. Halpern\altaffilmark{7},
G. Hinshaw\altaffilmark{2},
N. Jarosik\altaffilmark{3},
E. Komatsu\altaffilmark{8},
M. R. Nolta\altaffilmark{5},
N. Odegard\altaffilmark{2,9},
L. Page\altaffilmark{3},
D. N. Spergel\altaffilmark{6},
G. S. Tucker\altaffilmark{10},
J. L. Weiland\altaffilmark{2,9},
E. Wollack\altaffilmark{2},
E. L. Wright\altaffilmark{11}
}

\altaffiltext{1}{{\sl WMAP} is the result of a partnership between 
		 Princeton  University and NASA's Goddard Space Flight Center. 
		 Scientific guidance is provided by the 
		 {\sl WMAP} Science Team.}
\altaffiltext{2}{Code 665, Goddard Space Flight Center, Greenbelt, MD 20771}
\altaffiltext{3}{Dept. of Physics, Jadwin Hall, 
            Princeton University, Princeton, NJ 08544-0708}
\altaffiltext{4}{Dept. of Physics \& Astronomy, 
            The Johns Hopkins University, 3400 N. Charles St., 
	    Baltimore, MD  21218-2686}
\altaffiltext{5}{Canadian Institute for Theoretical Astrophysics, 
            60 St. George St, University of Toronto, 
	    Toronto, ON  Canada M5S 3H8}
\altaffiltext{6}{Dept. of Astrophysical Sciences, 
            Peyton Hall, Princeton University, Princeton, NJ 08544-1001}
\altaffiltext{7}{Dept. of Physics and Astronomy, University of
            British Columbia, Vancouver, BC  Canada V6T 1Z1}
\altaffiltext{8}{Univ. of Texas, Austin, Dept. of Astronomy, 
            2511 Speedway, RLM 15.306, Austin, TX 78712}
\altaffiltext{9}{ADNET Systems, %
	164 Rollins Ave. Suite 303, Rockville MD 20852}
\altaffiltext{10}{Dept. of Physics, Brown University, 
            182 Hope St., Providence, RI 02912-1843}
\altaffiltext{11}{PAB 3-909, UCLA Physics \& Astronomy, PO Box 951547, 
            Los Angeles, CA 90095--1547}

\email{Alan.J.Kogut@nasa.gov}

% --------------------- Abstract ---------------------
\begin{abstract}
We present a full-sky model of polarized Galactic microwave emission
based on three years of observations
by the Wilkinson Microwave Anisotropy Probe ({\sl WMAP})
at frequencies from 23 to 94 GHz.
The model compares maps of the Stokes $Q$ and $U$ components
from each of the 5 {\sl WMAP} frequency bands
in order to separate synchrotron from dust emission,
taking into account the spatial and frequency dependence
of the synchrotron and dust components.
This simple two-component model of the interstellar medium
accounts for at least 97\% of the polarized emission
in the {\sl WMAP} maps of the microwave sky.
Synchrotron emission dominates the polarized foregrounds
at frequencies below 50 GHz,
and is comparable to the dust contribution at 65 GHz.
The spectral index of the synchrotron component,
derived solely from polarization data,
is -3.2 averaged over the full sky,
with a modestly flatter index on the Galactic plane.
The synchrotron emission has 
mean polarization fraction 2--4\% in the Galactic plane
and rising to over 20\% at high latitude,
with prominent features such as the North Galactic Spur
more polarized than the diffuse component.
Thermal dust emission has polarization fraction
1\% near the Galactic center,
rising to 6\% at the anti-center.
Diffuse emission from high-latitude dust is also polarized
with mean fractional polarization
$0.036 \pm 0.011$.
\end{abstract}


% --------------------- Main text ---------------------

% \clearpage
% \newpage

\section{INTRODUCTION}

The Wilkinson Microwave Anisotropy Probe ({\sl WMAP})
has mapped the full sky
in the Stokes $I$, $Q$, and $U$ parameters
on angular scales $\theta > 0\fdg2$
in 5 frequency bands centered at 23, 33, 41, 61, and 94 GHz,
denoted K, Ka, Q, V, and W, respectively
\citep{bennett/etal:2003b, hinshaw/etal:prep}.
Polarized emission at microwave frequencies
is dominated by foreground emission from the Galaxy,
which at degree angular scales
is brighter than the cosmological signal
in all of the {\sl WMAP} bands
\citep{page/etal:prep}.
Extraction of cosmological information 
from the {\sl WMAP} polarization data
requires fitting for the astrophysical foregrounds.

Foreground polarization at microwave frequencies
is dominated by
a superposition of synchrotron and thermal dust emission.
Synchrotron emission
results from the acceleration of cosmic ray electrons
in the Galactic magnetic field.
For a power-law distribution of electron energies
$N(E) \propto E^{-p}$
propagating in a uniform magnetic field,
the resulting emission
is partially polarized
with fractional linear polarization
\begin{equation}
f_s = \frac{p+1}{p + 7/3}
\label{fs_vs_p_eq}
\end{equation}
aligned perpendicular to the magnetic field
\citep{rybicki/lightman:1979}.
The frequency dependence of synchrotron emission
is also related to the electron energy distribution,
$T(\nu) \propto \nu^\beta$
with spectral index
\begin{equation}
\beta = -\frac{p+3}{2}
\label{beta_vs_p_eq}
\end{equation}
where $T$ is in units of antenna temperature.
For spectral index $\beta \approx -3$
observed at microwave frequencies,
synchrotron emission
could have fractional polarization as high as $f_s \sim 0.75$.
Line-of-sight and beam averaging effects
will tend to reduce the observed polarization
by averaging over regions with different 
electron energy distribution
or magnetic field orientation.

Thermal dust emission can be partially polarized
as prolate dust grains
align with their long axis perpendicular to the magnetic field
\citep{davis/greenstein:1951}.
The dust emission efficiency
is greatest along the long axis,
leading to partial linear polarization
perpendicular to the magnetic field
(thus following the same direction as synchrotron emission).
The fractional polarization
depends on the grain size distribution
and is typically a few percent at millimeter wavelengths
\citep{Hildebrand:1988,
Hildebrand/etal:1999,
Vaillancourt:2002}.

Other sources of polarized emission are possible.
Extragalactic radio point sources,
although detected in the unpolarized temperature maps,
contribute negligibly to the observed polarization
on degree angular scales
\citep{page/etal:prep}.
Non-thermal emission from dust
via a population of small spinning grains
\citep{draine/lazarian:1998}
or partially magnetized grains
\citep{draine/lazarian:1999}
has been proposed to explain
the observed correlation
between far-infrared and microwave emission 
in the interstellar medium.
While emission from spinning dust 
is thought to be largely unpolarized,
emission from partially magnetized grains
could be substantially polarized 
provided the grains consist of a single magnetic domain
\citep{draine/lazarian:1999}.

\citet{page/etal:prep} present a simple model
of foreground polarization
based on the three-year {\sl WMAP} data.
Spatial templates for the synchrotron and dust emission,
fit to the high-latitude sky,
reduce the residual foregrounds 
in the corrected maps
to levels below the polarization
of the cosmic microwave background.
This paper compares the {\sl WMAP} polarization data
to the unpolarized synchrotron and dust emission
to model the polarized foregrounds,
including both the spatial and frequency variation
of the foregrounds
over the full sky.

%-----------------------------------------------------------
% Figure 1: Signal variance vs frequency
%-----------------------------------------------------------
\begin{figure}[t]
\plotone{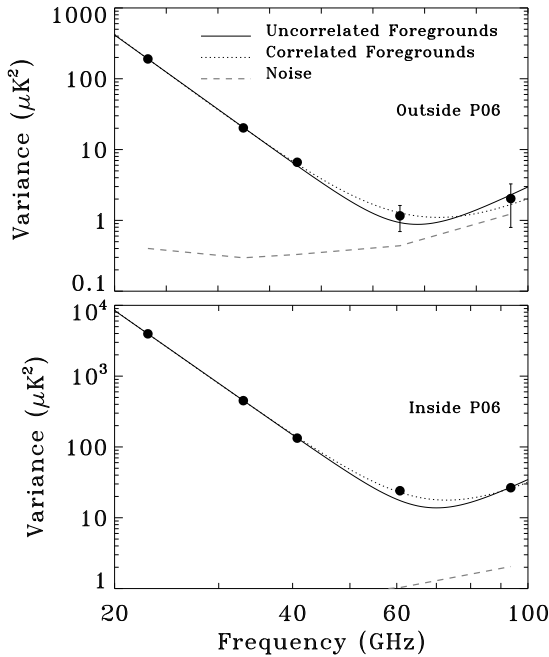}
\caption{Estimated signal variance
at each {\sl WMAP} frequency band
for 3\fdg7 pixels.
The solid line shows a two-component fit
with power-law synchrotron and thermal dust emission
assuming no spatial correlations between
the two components.
Allowing spatial correlations between the synchrotron and dust
better reproduces the observed spectra
(dotted lines).
The thick dashed line shows the 68\% confidence noise level.
Note the change in scale between panels:
the top panel shows high-latitude data outside the P06 sky mask,
while the bottom panel shows data inside P06 
dominated by emission from the Galactic plane.
\label{variance_fig} }
\end{figure}
%-----------------------------------------------------------

\section{FREQUENCY SPECTRUM}
The signal variance at each {\sl WMAP} frequency band
provides a simple, model-independent estimate 
of the frequency spectrum of polarized emission.
The data is a superposition of signal 
(either cosmic or foreground)
and noise,
$T = s + n$.
We estimate the signal variance
\begin{equation}
s^2 = \frac{
	\Sigma_{i\alpha}
	  w_{i \alpha}^a w_{i \alpha}^b T_{i \alpha }^a  T_{i \alpha }^b 
	   }{
	\Sigma_{i\alpha} w_{i \alpha}^a w_{i \alpha}^b
	   }
\label{variance_eq}
\end{equation}
where $i$ is a pixel index, 
$a$ and $b$ are map indices,
$\alpha$ denotes the Stokes $Q$ or $U$ maps,
and $w$ is a pixel weight.
We use different weights in different regions of the sky.
Pixels at low Galactic latitude
are signal-dominated and use unit weight,
$w_i = 1$.
At high latitudes, 
instrument noise becomes appreciable
and we use noise weight,
$w_i = N_i / \sigma_0^2$
where
$N_i$ is the number of observations in pixel $i$
and $\sigma_0$ is the noise per observation.
If the variance is computed from a single map
($a = b$),
instrument noise contributes a positive contribution
to each pixel,
$ \left< T^2 \right> = \left< (s + n)^2 \right> = s^2 + \left< n^2 \right>$.
We avoid this noise bias
by computing the cross variance
using maps from different years,
$a \ne b$.
Uncertainties in the noise model
then affect only the uncertainty in the estimated variance
but not the mean value itself.
We estimate the uncertainty in the computed cross-variance
using Monte Carlo simulations 
consisting of a model signal
plus noise realizations
derived from a Cholesky decomposition of the noise matrices
for each map
\citep{jarosik/etal:prep}.
The noise simulations thus include
the contribution of $1/f$ noise
as well as correlations between different pixels
and between the Stokes $Q$ and $U$ maps.
We add the same simulated foreground to each noise realization
prior to computing the variance,
to account for the additional scatter
from one realization to the next
caused by the signal-noise cross term
$2sn$.

Figure \ref{variance_fig} 
shows the frequency spectrum of the signal variance
for two regions of the sky
defined by the P06 polarization mask
\citep{page/etal:prep}:
the ``high signal'' region
inside the P06 mask,
dominated by the Galactic plane and North Galactic spur
where signals are clearly visible in the maps,
and a ``low signal'' region
outside the P06 mask
where noise is more prominent.
Both regions are dominated by a power-law component
$T \propto \nu^\beta$
with spectral index $\beta_s \approx -3$
consistent with synchrotron emission.
The rise in variance from V band to W band
(61 to 94 GHz)
requires a second component
with spectral index $\beta_d \approx 2$
consistent with thermal dust.
The observed variance is brighter than the expected cosmic signal
($s^2 \approx 0.14 ~\mu {\rm K}^2$ within 3\fdg7 ~pixels)
at all frequencies.
Detection of a cosmic signal
requires subtraction of foreground polarization
\citep{page/etal:prep}.

The observed steepening of the cosmic ray energy spectrum
is expected to steepen the synchrotron emission spectrum
relative to a pure power law
\citep{muller/tang:1987,
volk:1989}.
The polarization variance does not show the expected steepening 
above 30 GHz,
but rather shows a spectrum
flatter than a pure power law.
This could indicate either 
the presence of a third emission component
or non-zero spatial correlations 
between the synchrotron and dust components.
If the synchrotron and dust components
each follow a power law in frequency,
$T_{\rm synch} = S \nu^\beta_s$
and
$T_{\rm dust} = D \nu^\beta_d$,
then the variance becomes
\begin{equation}
T^2 = S^2 \nu^{2 \beta_s}   
    + D^2 \nu^{2 \beta_d} 
    + 2 r S D \nu^{\beta_s + \beta_d}
\label{2comp_eq}
\end{equation}
where $r$ is the spatial correlation
between the two components.
We fit the observed variance data
to a 2-component model
including spatial correlation,
and find a broad minimum with $r \sim 0.8$
(dotted lines in Figure \ref{variance_fig}).
Since the unpolarized synchrotron and dust foregrounds
tend to be highly correlated
\citep{bennett/etal:2003c,hinshaw/etal:prep},
it is not surprising to find that the correlation
persists in polarized emission from the same media.

\section{FOREGROUND POLARIZATION}
The {\sl WMAP} 3 year data 
map the polarized synchrotron emission
at high signal to noise ratio.
Polarized dust emission is less well constrained,
particularly at high Galactic latitudes.
We may use maps of the {\it unpolarized} dust
to constrain the maximum amplitude of polarized dust emission.
The dust polarization is
\begin{equation}
P_d(\hat{n},\nu) = f_d(\hat{n}) T_d(\hat{n}, \nu)
\label{dust_def}
\end{equation}
where
$P = ( Q^2 + U^2 )^{0.5}$
is the polarized amplitude,
$T_d$ is the unpolarized amplitude,
and
$f_d$ is the fractional polarization
in direction $\hat{n}$.
Since $f_d$ must lie in the range [0,1],
the polarized dust amplitude
can not exceed the total unpolarized emission in any pixel.

We thus model the polarized foregrounds
as a superposition of synchrotron and thermal dust emission,
\begin{eqnarray}
Q(\hat{n},\nu) &=& 
	P_s(\hat{n}) S(\hat{n},\nu)     \cos(2 \gamma_s(\hat{n})) \nonumber \\
 & + & f_d(\hat{n}) T_d(\hat{n}) D(\nu) \cos(2 \gamma_d(\hat{n})) \nonumber \\
U(\hat{n},\nu) &=& 
	P_s(\hat{n}) S(\hat{n},\nu)     \sin(2 \gamma_s(\hat{n})) \nonumber \\
 & + & f_d(\hat{n}) T_d(\hat{n}) D(\nu) \sin(2 \gamma_d(\hat{n})) 
\label{model_eq}
\end{eqnarray}
where
$P_s(\hat{n})$ is the (polarized) synchrotron amplitude.
We model the frequency dependence of the dust 
as a spatially invariant power law in antenna temperature,
\begin{equation}
D(\nu) = \left( 
\frac{\nu}{\nu_0}
\right)^{\beta_d}
\label{xi_dust}
\end{equation}
with spectral index $\beta_d = 2$.
Synchrotron emission may be approximated as a power law,
but the spectral index varies with position on the sky
and may steepen with frequency.
We thus model the synchrotron frequency dependence as
\begin{equation}
S(\nu, \hat{n}) = 
\left( \frac{\nu}{\nu_0} \right)^{ {\beta_s(\hat{n}) + C \log (\nu / \nu_0 )} }
\label{xi_synch}
\end{equation}
allowing the spectral index $\beta_s(\hat{n})$
to vary with position.
The model includes a logarithmic steepening of the spectrum with frequency,
with $C \approx -0.4$ typical of the synchrotron emission
in the {\sl WMAP} unpolarized temperature data
\citep{hinshaw/etal:prep,bennett/etal:2003c}.

The polarization angle
$\gamma(\hat{n}) = 0.5 \arctan( U(\hat{n}) / Q(\hat{n}) ) $
is defined with respect to the Galactic meridian.
Synchrotron emission dominates the lowest frequency channels,
providing a template
for the synchrotron polarization angle $\gamma_s(\hat{n})$.
We use the 22 GHz (K-band) data alone
to compute the synchrotron polarization angle $\gamma_s$.
In principle, this contains a small level of contamination
from the cosmic signal.
An alternative method computes the polarization angle
from the K-Ka or K-Q difference maps.
This eliminates any cosmic signal
at the cost of increased noise.
We use Monte Carlo simulations to compare the two methods.
For noise levels typical of the {\sl WMAP} three-year polarization maps,
direct fitting to the K-band data
better reproduces the input pattern used to generate the simulations.

In this paper,
we do not assume that the dust follows the same polarization direction
as the synchrotron,
so that the synchrotron polarization angle $\gamma_s(\hat{n})$ 
may differ from the dust polarization angle $\gamma_d(\hat{n})$.
This allows for dust emission to originate 
from a different spatial distribution
compared to synchrotron emission
along each line of sight.
We adopt the dust polarization angles
from the starlight model described in
\citet{page/etal:prep},
and compare this model
to one in which 
$\gamma_d = \gamma_s$ everywhere on the sky.

%-----------------------------------------------------------
% Figure 2: Map of synchrotron spectral index
%-----------------------------------------------------------
\begin{figure}
\includegraphics[width=3.25in]{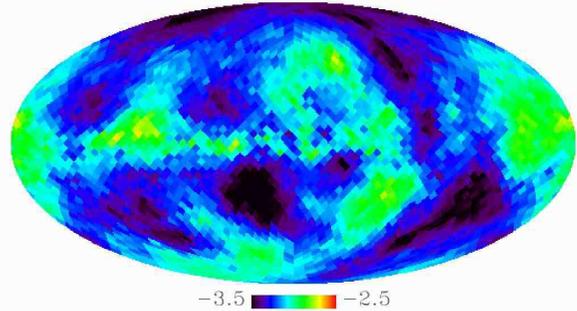}
\caption{Spectral index of polarized synchrotron emission
between 22 and 33 GHz
(Mollweide projection in Galactic coordinates).
The map has been smoothed with a position-dependent tophat
varying from 7\arcdeg ~radius near the plane
to 18\arcdeg ~radius at high latitude.
\label{beta_map} }
\end{figure}
%-----------------------------------------------------------

The foreground model does not assume that the 
synchrotron spectral index is identical along
all lines of sight,
but fits a map of the spectral index
as described in $\S$\ref{synch_model}.
Given the polarization angles 
$\gamma_s(\hat{n})$ and $\gamma_d(\hat{n})$
and the map of the synchrotron spectral index
$\beta_s(\hat{n})$,
we fit the {\sl WMAP} 3-year band-averaged
polarization maps
to derive the polarized synchrotron amplitude $P_s(\hat{n})$,
synchrotron spectral correction $C$,
and
the dust fractional polarization $f_d(\hat{n})$
(Eq. \ref{model_eq}).
Based on signal to noise,
we fit these parameters simultaneously
but on different angular scales.
Synchrotron emission dominates the low-frequency data.
We fit for the synchrotron amplitude
$P_s(\hat{n})$ (normalized to K band)
in each of 3072 equal area pixels
at HEALPix resolution 4
or $N_{\rm side} = 16$
\citep{gorski/etal:2005}.
Polarized dust emission is considerably fainter than synchrotron
in the {\sl WMAP} frequency bands.
We thus fit for the dust fractional polarization 
$f_d(\hat{n})$
in 6 coarse sky regions,
holding $f_d$ constant within each region
while simultaneously fitting
for the synchrotron amplitude
in each of the smaller HEALPix pixels.
We fit for the synchrotron spectral correction $C$
in two regions defined by the P06 polarization mask:
one value for the high-latitude region outside the mask
and an independent value for the low-latitude region inside the mask.
We model the unpolarized dust emission $T_d(\hat{n}, \nu)$
using the \citet{finkbeiner/davis/schlegel:1999}
(FDS) model 8,
evaluated at 94 GHz
and
scaled with uniform spectral index $\beta_d = 2$.
The results do not depend sensitively
on the choice for $\beta_d$.

\subsection{Polarized Synchrotron Model
\label{synch_model}}

Synchrotron emission may be approximated as a power law in antenna temperature,
with spectral index $\beta_s(\hat{n})$ varying with position across the sky.
\citet{hinshaw/etal:prep} present a maximum-entropy model 
of unpolarized synchrotron emission
derived from {\sl WMAP} data.
We may use this maximum-entropy model 
to specify the frequency dependence of the polarized synchrotron emission,
including both the spatial variation 
and the steepening of the spectrum 
with respect to a pure power law.
The unpolarized maximum-entropy model, however,
suffers from confusion with free-free emission,
particularly at low Galactic latitudes.

%-----------------------------------------------------------
% Figure 3: Synchrotron spectral index vs latitude
%-----------------------------------------------------------
\begin{figure}[t]
\includegraphics[angle=90,width=3.25in]{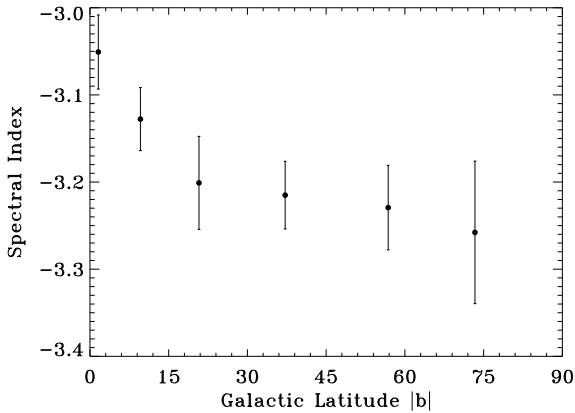}
\caption{Spectral index of polarized synchrotron emission
binned by Galactic latitude.
The spectral index steepens off the Galactic plane.
Similar behavior is observed in external edge-on spiral galaxies.
\label{beta_vs_lat} }
\end{figure}
%-----------------------------------------------------------

Synchrotron emission dominates the polarization data 
at 22 and 33 GHz
(K and Ka bands),
with negligible contribution from thermal dust
(Fig. \ref{variance_fig}).
We thus use the {\sl WMAP} polarization data at K and Ka bands
to map the synchrotron spectral index $\beta_s(\hat{n})$.
We fit for the spectral index
in each of 3072 pixels,
imposing a uniform prior
$-4.0 < \beta_s < -2.0$
to reduce the effects of instrument noise.
The resulting map is still noise dominated,
particularly at high latitude.
We further reduce the noise contribution
by smoothing the spectral index map
using a tophat window function.
We select the tophat radius
and verify the accuracy of the fitting procedure
using Monte Carlo realizations
consisting of a fixed Galactic signal
added to random realizations of instrument noise.
At each pixel,
we smooth the spectral index maps from each realization
using a progressively larger tophat radius,
until the standard deviation of the ensemble of smoothed realizations,
evaluated at that pixel,
falls below a threshold 
$\delta \beta_s(\hat{n}) < 0.1$.
The resulting smoothing radius
varies across the sky
from 7\arcdeg ~near the plane
to 18\arcdeg ~at high latitude.

We investigate the dependence of the recovered spectral index map
on the prior imposed during the pixel-by-pixel fit.
Changing the prior to a wider range
$-4.5 < \beta_s < -1.5$
or a narrower range
$-3.5 < \beta_s < -2.5$
produces a systematic shift
$\delta \beta_s \approx 0.1$
in the mean of the smoothed spectral index map,
comparable to random uncertainty,
but does not significantly alter the structure 
on smaller angular scales.
Our choice of prior is informed by Monte Carlo realizations,
where
the ensemble average of the smoothed realizations
accurately recovers the noiseless spectral index map
used to generate the simulations.

Figure \ref{beta_map}
shows the polarized synchrotron spectral index
from the {\sl WMAP} 3-year data.
The mean value averaged over the full sky
is $\langle \beta_s \rangle = -3.2$.
The spectral index is flatter along the Galactic plane
with steeper values at higher latitude.
Figure \ref{beta_vs_lat}
shows the polarization spectral index
binned by Galactic latitude.

The smoothing applied to the spectral index maps
creates spatial correlations in the smoothed maps.
We account for this in Figure \ref{beta_vs_lat} 
by plotting the effective uncertainty
in the binned data,
$\delta \beta_s^{\rm eff} = \sigma_i / \sqrt{N_i^{\rm eff}}$,
where
$\sigma_i$ is the standard deviation of the smoothed values
from all pixels
in the $i^{\rm th}$ latitude bin,
and
$N_i^{\rm eff}$ is the number
of independent spatial regions within that bin.
The spectral index
steepens from 
$\beta_s \approx -3.05$ along the plane
to 
$\beta_s \approx -3.25$ at latitude $|b| > 30\arcdeg$.
Polarization data
provide a measurement of the synchrotron spectral index
independent of the temperature data
and not subject to confusion from competing emission mechanisms.

%-----------------------------------------------------------
% Figure 4: Map of synchrotron model and fractional polarization
%-----------------------------------------------------------
\begin{figure}[b]
\plotone{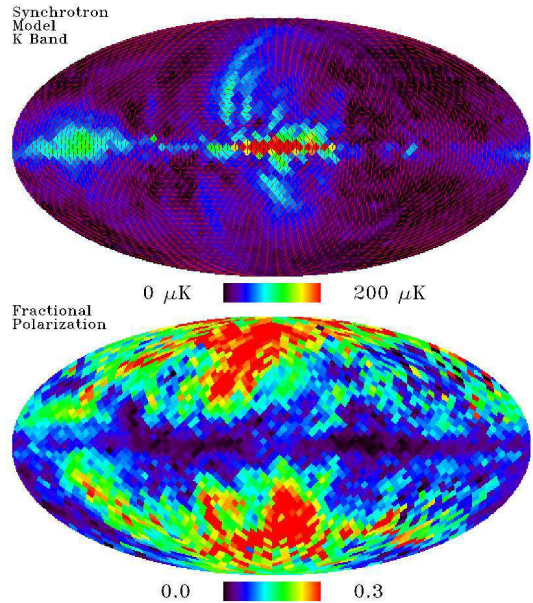}
\caption{Polarized synchrotron model parameters.
(Top) Antenna temperature $P = (Q^2 + U^2)^{0.5}$
of polarized synchrotron emission 
at K band.
Vectors indicate the polarization direction.
(Bottom) Synchrotron fractional polarization,
derived by dividing the polarized model
by the unpolarized maximum-entropy synchrotron model
at K band.
The polarization fraction is 3--5\% on the Galactic plane,
increasing above 20\% within the North Galactic Spur
and its southern extension.
\label{fs_map} }
\end{figure}
%-----------------------------------------------------------

%-----------------------------------------------------------
% Figure 5: Histogram of synchrotron fractional polarization
%-----------------------------------------------------------
\begin{figure}[t]
\plotone{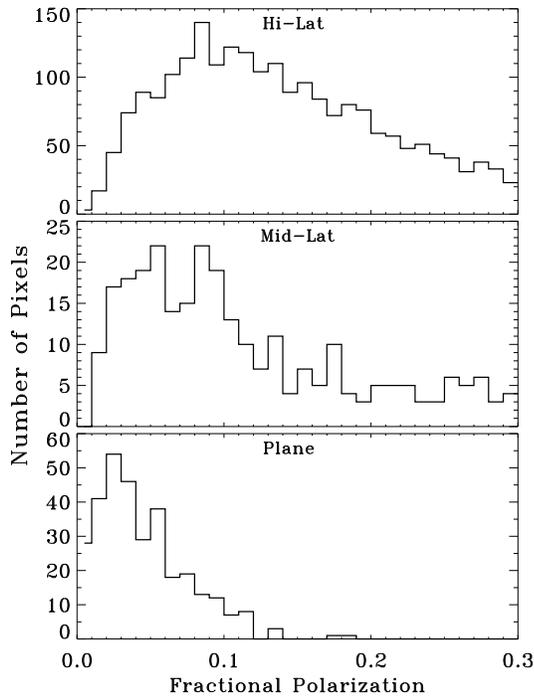}
\caption{Histogram of synchrotron fractional polarization $f_s$.
(Top) High latitude region outside the P06 mask.
(Middle) Mid-latitude region inside P06 but with $|b| > 5\arcdeg$.
(Bottom) Galactic plane, $|b| < 5\arcdeg$.
\label{fs_hist} }
\end{figure}
%-----------------------------------------------------------

Figure \ref{fs_map} shows the model polarized synchrotron emission.
The polarized synchrotron amplitude $P_s(\hat{n})$
is dominated by the lowest frequency data
and is nearly identical to the K-band map.
We derive the synchrotron fractional polarization
by comparing the K-band model amplitude
to the unpolarized K-band maximum-entropy synchrotron model
\citep{hinshaw/etal:prep},
assuming that this model accurately represents the unpolarized
synchrotron amplitude.
Although the Galactic plane dominates the polarized emission in the maps,
the corresponding fractional polarization is low,
indicative of contributions from multiple emitting regions
along the line of sight.
The North Galactic Spur and its southern extension are more prominent
in polarization than in intensity,
indicating localized enhancements in the polarization fraction.
Figure \ref{fs_hist} shows a histogram of the 
synchrotron fractional polarization
for several regions of the sky.
The Galactic plane ($|b| < 5\arcdeg$)
is less than 10\% polarized,
with mode $f_s = 0.02$
and mean $f_s = 0.05$.
The North Galactic Spur has enhanced fractional polarization,
$f_s \approx 0.3$,
while the high-latitude sky outside the P06 mask
has mode $f_s = 0.14$ and mean $f_s = 0.15$.

\begin{figure}[b]
\includegraphics[angle=90,width=3.25in]{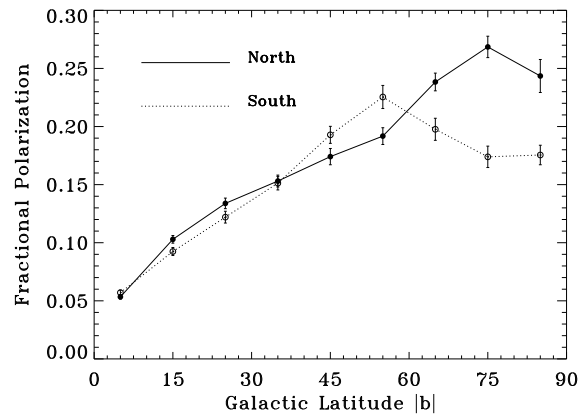}
\caption{Synchrotron fractional polarization $f_s$
binned by Galactic latitude
for the northern (filled symbols)
and southern (open symbols) hemispheres.
Each point shows the mean fractional polarization $f_s$ 
for all pixels in a latitude bin
with signal-to-noise ratio greater than one.
The error bars show the uncertainty in the mean
based on the scatter of values
within each latitude bin.
\label{synch_vs_lat} }
\end{figure}

Figure \ref{synch_vs_lat}
shows the fractional polarization
binned by Galactic latitude.
Within narrow latitude bins,
the distribution of observed fractional polarization values
approximates a Gaussian distribution.
Figure \ref{synch_vs_lat} shows
the mean $ \langle f_i \rangle$ within the $i^{\rm th}$ bin
with uncertainty
$\delta f_i = \sigma_i / \sqrt{N_i}$
defined by the standard deviation $\sigma_i$ 
of the $N_i$ pixels within that bin.
The uncertainties thus reflect the precision
to which the mean may be defined within each latitude bin,
which average over a larger range of fractional polarization
({\it cf} Figure \ref{fs_hist}).
The mean fractional polarization rises linearly
from 4\% near the plane to 20\% at latitude $|b| > 50\arcdeg$.
The north polar cap region is more polarized than the south polar cap,
although this may be an artifact of the unpolarized synchrotron model
used to define the synchrotron fractional polarization.
Both the low polarization in the plane of the Galaxy
and the trend toward higher fractional polarization
at high latitude
are similar to previous measurements 
of external spiral galaxies.
Edge-on spirals typically exhibit 1--3\% polarization in the disk,
rising linearly to 10--20\% at distances of several kpc 
from the major axis
\citep{
hummel/beck/dahlem:1991,
sukumar/allen:1991,
dumke/etal:1995,
dumke/krause:1998, 
dumke/krause/wielebinski:2000}.
The large-scale polarization of the Milky Way
is thus unremarkable compared to other galaxies.

The foreground model
directly fits the polarized synchrotron amplitude $P_s(\hat{n})$
and compares this fitted amplitude
to a model of the unpolarized synchrotron emission
to estimate the fractional polarization $f_s$.
A recent comparison of the unpolarized {\sl WMAP} K-band map at 22 GHz
with a full-sky map at 19 GHz
suggests that much of the emission near the Galactic plane
results from
a component spatially correlated with thermal dust emission
but with with spectral index $\beta \approx -1.7$,
consistent with spinning dust
\citep{boughn/pober:2006}.
To the extent that the maximum-entropy synchrotron model
over-estimates the unpolarized synchrotron amplitude
by ignoring spinning dust,
the fractional polarization $f_s$
derived above
becomes a lower limit
to the true synchrotron fractional polarization.
Since the values for $f_s$ derived using the maximum-entropy model
agree with observations of external edge-on spiral galaxies
at much lower frequencies
(where spinning dust is negligible),
requiring a substantial fraction of the unpolarized emission at 22 GHz
to originate from spinning dust
would require some compensating depolarization
of the synchrotron component
({\it e.g.} line-of-sight effects
or a tangled magnetic field)
to keep the Milky Way in family with similar external galaxies.

%-----------------------------------------------------------
% Figure 6: Synchrotron fractional polarization vs latitude
%-----------------------------------------------------------
%-----------------------------------------------------------

\subsection{Polarized Dust Model}

Polarized dust emission is brighter than synchrotron
only in the highest frequency channel,
and only significantly so on the Galactic plane.
Since the dust fractional polarization $f_d$ 
is by definition positive,
noise in the W band data
can create a positive bias in the fitted dust solution.
We estimate this bias using Monte Carlo simulations,
comparing the recovered synchrotron and dust parameters
to known inputs in each coarse sky region
for 300 realizations
consisting of synchrotron, dust, CMB, 
and instrument noise 
generated using Cholesky decomposition of the full noise matrix
in each DA
\citep{jarosik/etal:prep}.
The recovered dust solution
is nearly identical to the input,
with bias less than 10\% of the statistical uncertainty.
We do not apply a correction for this effect.

% -------------- Table 1: Dust fractional polarization --------------
\begin{table}[t]
\begin{center}
\caption{Dust Fractional Polarization}
\label{fd_table}
\begin{tabular}{l c c c}
\tableline 
\tableline 
Region  & Galactic & Galactic  & Fractional \\
        & Latitude & Longitude & Polarization \\
	& (deg)    &  (deg)    &   $f_d$ \\
\tableline 
1  &  -10 to  10  &  -135 to  -45  		  &  0.010 $\pm$ 0.003	\\
2  &  -10 to  10  &   -45 to   45  		  &  0.010 $\pm$ 0.005	\\
3  &  -10 to  10  &    45 to  135  		  &  0.012 $\pm$ 0.005	\\
4  &  -10 to  10  &   135 to -135  		  &  0.064 $\pm$ 0.013	\\
5  &   \multicolumn{2}{c}{$|b| >$ 10, inside P06} &  0.020 $\pm$ 0.014	\\
6  &   \multicolumn{2}{c}{Outside P06}		  &  0.036 $\pm$ 0.011	\\
\tableline
\end{tabular}
\end{center}
\end{table}
%----------------------------------------------------------------------------

Table \ref{fd_table} shows the fractional polarization
of the dust in each coarse sky region.
The fractional polarization in the plane increases
away from the Galactic center,
with $f_d = 0.01$ at the Galactic center
rising to 0.06 at the anti-center.
The fractional polarization increases with latitude,
reaching a value 
$f_d = 0.036 \pm 0.011$
outside the P06 polarization mask.

Equation \ref{model_eq} fits a single parameter
to the high-latitude dust 
(outside the P06 mask).
We have repeated the fit
using more ``coarse'' sky regions outside the P06 mask.
The faint amplitude of the dust signal
combined with the lack of small-scale structure
in the high-latitude template map $D(\hat{n})$
prevent a statistically significant determination of
substructure in the high-latitude polarized dust signal.

\section{DISCUSSION}

\begin{figure*}[t]
\plotone{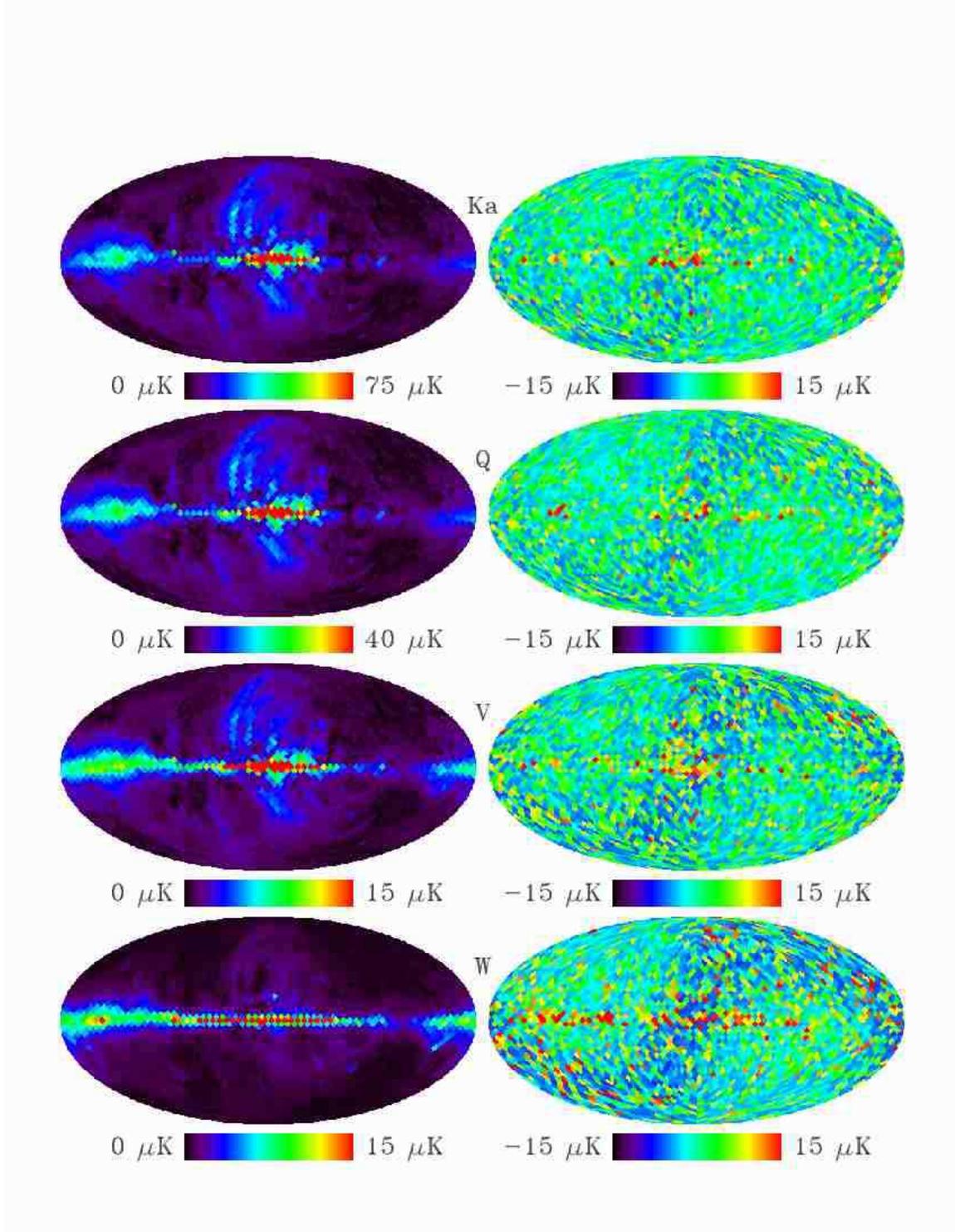}
\caption{Polarized foreground model and residuals 
for the pixel-by-pixel polarization model.
(Left) Antenna temperature of polarized emission 
at each band.
The polarization direction is not shown.
(Right) Residuals after subtracting foreground model.
For visualization, the residuals $P = \sqrt{Q^2 + U^2}$
are shown, corrected for the positive noise bias
using Monte Carlo simulations.
The model and $\chi^2$ are 
computed directly from the Stokes $Q$ and $U$ maps.
\label{resid_fig} }
\end{figure*}

We model the polarized foreground emission
in each pixel
as a superposition of synchrotron emission
with spatially variable spectral index 
determined from the {\sl WMAP} polarization data
plus dust emission
traced by FDS dust model
scaled by a spatially variable fractional polarization.
Figure \ref{resid_fig} shows the model 
and residuals for each frequency band.
A simple test compares the variance of the foreground-cleaned maps
to the variance of the uncleaned maps
(Equation \ref{variance_eq}).
The foreground model removes over 97\% of the power
at Ka and Q bands where foregrounds are brightest.
Outside the P06 mask,
the variance of the cleaned maps
is close to the value expected for instrument noise alone.

Table \ref{chisq_table} 
compares the template model
described in \citet{page/etal:prep}
to the pixel model described in this paper,
using goodness-of-fit statistic
\begin{equation}
\chi^2 = \Sigma_{ij\alpha\beta} R_{i\alpha} N_{ij\alpha\beta}^{-1} R_{j\beta}
\label{chisq_eq}
\end{equation}
where
$N^{-1}$ is the inverse noise matrix
and
$R_{i\alpha}$ is the (data-model) residual
in pixel $i$ and Stokes parameter $\alpha$.
The synchrotron amplitude $P_s$ and polarization angle $\gamma_s$
in the pixel model
are derived almost entirely from the K-band data,
leaving nearly zero residuals at K band.
Since the synchrotron amplitude and polarization angle
may be treated as an external (K-band) template,
we assess the goodness of fit 
using only the 4 higher-frequency bands,
counting only the parameters fitted using data at these bands.
The high-latitude region outside the P06 mask
contains 2267 pixels 
in each of the Stokes $Q$ and $U$ maps
in each of the 4 frequency bands.
The synchrotron spectral index map
has 36 independent regions outside the P06 mask
(after accounting for smoothing),
requiring 36 parameters in the extended model.
We add one additional parameter for 
the synchrotron spectral correction $C$
plus another for
the dust fractional polarization $f_d$
to obtain a total of 38 parameters fitted to 18136 data points
outside the P06 mask.
For the 805 pixels inside the P06 mask we fit 
31 parameters for the synchrotron spectral index,
5 parameters for dust,
plus one parameter for the synchrotron spectral correction
for a total of 
36 parameters fitted to 6440 low-latitude data points.
The template model,
by comparison,
fits only 2 parameters to each band,
one each for the dust and synchrotron template maps.

% -------------- Table 2: Residual chi-squared --------------
\begin{table}[b]
\begin{center}
\caption{$\chi^2$ of Residual Maps}
\label{chisq_table}
\begin{tabular}{l c c c c}
\tableline 
\tableline 
Model	        &  Band &  $\chi^2$	&   DOF		&  $\chi^2$/DOF \\
\tableline 
Pixel Model	&  Ka	&   4604	&   4524	&  1.018  \\
Outside P06 	&  Q	&   4521	&   4524	&  1.000  \\
	 	&  V	&   4600	&   4524	&  1.017  \\
	 	&  W	&   4774	&   4524	&  1.055  \\
	 	& All   &  18505	&  18098	&  1.023  \\
\tableline 
Pixel Model 	&  Ka	&   1626	&   1601	&  1.016  \\
Inside P06 	&  Q	&   1541	&   1601     	&  0.963  \\
	 	&  V	&   1670	&   1601     	&  1.043  \\
	 	&  W	&   1773	&   1601     	&  1.107  \\
	 	& All   &   6611	&   6404	&  1.032  \\
\tableline
Template Model 	&  Ka	&   4968	&   4532	&  1.096  \\
Outside P06 	&  Q	&   4616	&   4532	&  1.019  \\
	 	&  V	&   4608	&   4532	&  1.017  \\
	 	&  W	&   4760	&   4532	&  1.050  \\
	 	& All   &  18952	&  18128	&  1.045  \\
\tableline 
Template Model	&  Ka	&   2403	&   1608	&  1.494  \\
Inside P06 	&  Q	&   2081	&   1608     	&  1.294  \\
	 	&  V	&   2704	&   1608     	&  1.682  \\
	 	&  W	&   4946	&   1608     	&  3.076  \\
	 	& All   &  12034	&   6432	&  1.871  \\
\tableline
\end{tabular}
\end{center}
\end{table}
%----------------------------------------------------------------------------

Fitting a spatially variable synchrotron spectral index
provides a statistically significant improvement
($\Delta \chi^2 = 447$ for 30 additional model parameters
outside the P06 mask)
compared to the template model,
which assumes the same spatial pattern for the synchrotron
at all bands.
Most of this improvement
comes at Ka and Q bands where synchrotron is brightest.
Although the pixel-based model
has a lower $\chi^2$ at V band,
the difference at this band alone
is not statistically significant.
The template model has slightly better $\chi^2$ at W band
outside the P06 mask.
Not surprisingly, the pixel-based model provides a better fit
at all bands inside the P06 mask.

%-----------------------------------------------------------
% Figure 7: Map of model residuals
%-----------------------------------------------------------
%-----------------------------------------------------------

The pixel-based model
finds a synchrotron spectrum 
marginally flatter than a power law,
with best-fit spectral correction
$C = 0.3 \pm 0.1$.
This result is unchanged with
modifications of the model
(polarization angle, 
synchrotron spectral index map,
dust spectral index)
and is independent of the sky cut.
The apparent flattening is a robust feature of the data,
derived independently from analyses
of the data variance ($\S$2),
the pixel-based model ($\S$3),
and the template model
\citep{page/etal:prep}.
We obtain the same value for $C$
fitted independently
to the regions inside and outside the P06 mask.
We note that the value for $C$ 
is heavily weighted by the data at $Q$ and $V$ bands
where both the synchrotron and dust are faint.
Forcing $C=0$ to fit a pure power law
increases the $\chi^2$ by 10 outside the P06 mask,
while a steepening spectrum
($C = -0.4)$ increases $\chi^2$ by 45.
In both cases the the fitted dust parameters shift slightly
compared to the best-fit model.
We define a systematic uncertainty
for the dust fractional polarization
as the change in $f_d$ 
induced as the
synchrotron spectral correction parameter $C$
varies from 0.3 to -0.4.
The uncertainties reported in Table \ref{fd_table}
include this systematic uncertainty 
added in quadrature with the statistical uncertainty.

The $\chi^2$ analysis quantifies the dependence 
of the synchrotron spectral index map
on the choice of prior used in the fitting process
($\S$3.1). 
We repeat the full foreground analysis,
replacing the spectral index map $\beta_s(\hat{n})$
in Eqs. \ref{model_eq} -- \ref{xi_synch}
with a new map
derived using a different prior
before re-computing the best-fit
values for the foreground model.
We find a broad minimum in $\chi^2$
near prior $-4.0 < \beta_s < -2.0$,
consistent with the Monte Carlo tests
described in $\S$3.1.
Changing to a more restrictive prior
$-3.5 < \beta_s < -2.5$ increases $\chi^2$ by 33,
while the looser prior
$-4.5 < \beta_s < -1.5$ increases $\chi^2$ by 2.

A $\chi^2$ test confirms that the structure
in the synchrotron spectral index map
is dominated by a steepening of the spectral index
off the Galactic plane.
We create a map of the synchrotron spectral index
in which the value at each pixel
depends only on Galactic latitude
(first 4 bins from Figure \ref{beta_vs_lat}).
We then compare the foreground model
derived using this spectral index map
to a model derived using
a uniform spectral index
equal to the mean of the best-fit spectral index map.
Allowing the spectral index to vary with latitude
near the plane
improves the $\chi^2$ by 88 using only 4 additional parameters.

We also test the assumption that the polarized dust distribution
may be traced by the polarization angle derived from 
observations of polarized starlight.
We repeat the fit in Eq. \ref{model_eq},
replacing the dust polarization angle $\gamma_d$
with the synchrotron polarization angle $\gamma_s$
derived from the {\sl WMAP} K-band data.
The fit is marginally better ($\Delta \chi^2 = 9$)
when the polarized dust emission follows the magnetic field
inferred from polarized starlight
instead of the magnetic field derived from the synchrotron component.

We may also use the polarization data
to constrain emission from 
any additional component such as spinning dust.
Direct comparison of the polarized maps
yields spectral index $\beta _s \approx -3.05$
near the Galactic plane,
consistent with synchrotron emission
but inconsistent with 
the inverted spectrum
found by \citet{boughn/pober:2006}.
We test for polarized emission
traced by the dust morphology
by modifying Eq. \ref{model_eq}
to include additional terms
\begin{eqnarray}
Q_{\rm spin}(\hat{n},\nu) &=& 
	f_{\rm spin}(\hat{n}) T_d(\hat{n}) D_{\rm spin}(\nu) \cos(2 \gamma_d(\hat{n})) \nonumber \\	
U_{\rm spin}(\hat{n},\nu) &=& 
	f_{\rm spin}(\hat{n}) T_d(\hat{n}) D_{\rm spin}(\nu) \sin(2 \gamma_d(\hat{n})) 	
\label{spin_eq}
\end{eqnarray}
where
$f_{\rm spin}$ is the fractional polarization 
of the spinning dust component
with frequency dependence
$D_{\rm spin}(\nu)$
peaking near 20 GHz
\citep{draine/lazarian:1998}.
We repeat the fit
using either
a spatially invariant synchrotron spectral index
or
the spatially variable map of polarized synchrotron emission
derived from the K-Ka band comparison
(Fig. \ref{beta_map}).
We fit the spinning dust term
within the same coarse pixels as the thermal dust emission,
but make no assumption
concerning the relative amplitudes of the
thermal vs spinning dust components.
In all cases,
the fitted values are consistent with
no spinning dust contribution to the polarization data,
$f_{\rm spin} = 0$.
Spinning dust contributes less than 1 percent
of the observed polarization signal variance
in any band.

\citet{page/etal:prep}
analyzed the {\sl WMAP} polarization data outside the P06 mask
to detect a cosmic signal
at amplitude
$\ell (\ell + 1) C_{\ell}^{EE} / 2\pi 
= 0.086 \pm 0.029 ~\mu{\rm K}^2$
averaged over multipole moments $\ell$ = 2 to 6.
Their analysis used a foreground model
derived by fitting template maps
to the high-latitude portion of the {\sl WMAP}
Stokes $Q$ and $U$ data.
This paper extends the foreground model
to include
spatial variation and frequency dependence
of the synchrotron spectral index
along with
spatial variation in the dust fractional polarization.
The extended foreground model
allows an improved understanding 
of the astrophysical foregrounds
but does not significantly alter the
cosmological conclusions.
Fitting the variance of the residual maps
as a superposition of instrument noise
plus a cosmological signal
yields a best-fit value
$s^2 = 0.38 \pm 0.21 ~\mu{\rm K}^2$
for the CMB variance
in thermodynamic temperature units,
compared to the expected value
$s^2 = 0.14 ~\mu{\rm K}^2$
derived from the \citet{page/etal:prep} detection.

%-----------------------------------------------------------
% Figure 8: Cleaned power spectra
%-----------------------------------------------------------
\begin{figure}[t]
\plotone{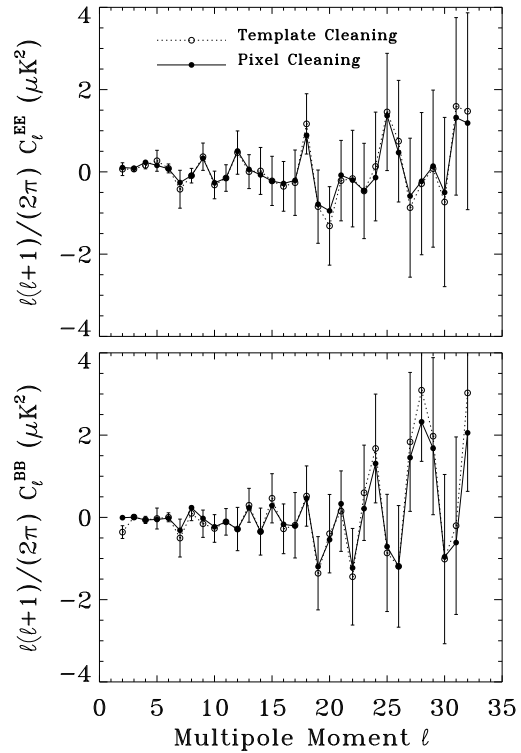}
\caption{CMB power spectra outside the P06 mask,
derived from
Q and V band data
after subtracting two different foreground models.
(top) EE power spectra.
(bottom) BB power spectra.
The CMB spectra are robust with respect
to different foreground models.
\label{cleaned_cl_fig} }
\end{figure}
%-----------------------------------------------------------

The cross variance analysis
provides a simple, model-independent limit
to residual signals in the cleaned maps.
Errors in the instrument noise model
affect the quoted uncertainty 
but do not bias the fitted value.
However, the variance analysis does not account
for spatial correlations in the instrument noise.
We have repeated the likelihood analysis
by correcting the Q and V band polarization maps
using the pixel-based foreground model
instead of the template model.
Figure \ref{cleaned_cl_fig}
shows the power spectra of the CMB signal
outside the P06 mask,
derived from the weighted sum
of the cleaned cross spectra at Q and V bands
on angular scales up to the pixel size
({\it cf} Fig 21 of \citet{page/etal:prep}).
The two cleaning techniques produce nearly identical results;
the only change larger than the noise uncertainty
is for the BB power spectrum at multipole moment $\ell=2$.
We quantify the effect of different foreground cleaning
on cosmological parameter estimation
by jointly fitting the EE and TE data
using the stand-alone likelihood code
without foreground marginalization
as described in \citet{page/etal:prep} Table 9.
Changing the foreground model
moves the fitted value for the optical depth
from
$\tau = 0.092 ^{+0.029}_{-0.030}$
to
$\tau = 0.098$,
a shift of roughly a fifth of the uncertainty.
Systematic uncertainties in cosmological parameters
caused by the foreground polarization
are small compared to the noise levels
of the {\sl WMAP} 3-year data.

\section{CONCLUSIONS}

{\sl WMAP} detects polarized emission over the full microwave sky.
The frequency dependence of this emission 
is consistent with a superposition of 
polarized synchrotron and thermal dust emission.
A pixel-by-pixel fit to the Stokes $Q$ and $U$ maps
from the band-averaged 3-year data
separates the emission
into synchrotron and dust components.
Synchrotron emission dominates the polarization
below frequencies of about 60 GHz.
The spectral index of the synchrotron component,
derived solely from polarization data,
is -3.2 averaged over the full sky,
with a modestly flatter index on the Galactic plane.
Comparison of the polarized synchrotron emission
to a maximum-entropy model of the unpolarized emission
shows typical fractional polarization
of 2--4\% near the Galactic plane,
rising to  20\% at latitude $|b| > 50\arcdeg$.
Prominent structures such as the North Galactic Spur
appear as localized enhancements in the polarization fraction.
We detect polarized dust emission 
with fractional polarization 
increasing from a minimum of 1\% near the Galactic center
to 6\% at the anti-center.
Diffuse emission from high-latitude dust is also polarized
with mean fractional polarization
$0.036 \pm 0.011$.
This simple two-component model of the interstellar medium
accounts for at least 97\% of the polarized emission
in the {\sl WMAP} maps of the microwave sky.

\acknowledgements

The {\sl WMAP} mission is made possible by the support of the 
Science Mission Directorate at NASA Headquarters 
and by the hard and capable work of scores of 
scientists, engineers, technicians, machinists, 
data analysts, budget analysts, 
managers, administrative staff, and reviewers. 
This research has additionally been supported
under NASA LTSA03-000-0090
and ATP award NNG04GK55G.
We acknowledge use of the HEALPix software package.

% --------------------- References --------------------------------
% Results from BibTex manually stuffed into manuscript LaTeX file
%------------------------------------------------------------------

% Exit, stage left
\end{document}